# A comparison of Gap statistic definitions with and without logarithm function

Mojgan Mohajer†

*Helmholtz Zentrum München and Department of Statistics, University of Munich*

Karl-Hans Englmeier

*Helmholtz Zentrum München*

Volker J. Schmid

*Department of Statistics, University of Munich*

**Summary**. The Gap statistic is a standard method for determining the number of clusters in a set of data. The Gap statistic standardizes the graph of $\log(W_k)$, where $W_k$ is the within-cluster dispersion, by comparing it to its expectation under an appropriate null reference distribution of the data. We suggest to use $W_k$ instead of $\log(W_k)$, and to compare it to the expectation of $W_k$ under a null reference distribution. In fact, whenever a number fulfills the original Gap statistic inequality, this number also fulfills the inequality of a Gap statistic using $W_k$, but not *vice versa*. The two definitions of the Gap function are evaluated on several simulated data sets and on a real data of DCE-MR images.

## 1. Introduction

In clustering methods the number of clusters is either a direct parameter, or it may be controlled by other parameters of the method. Estimating the proper number of clusters is an important problem in selecting the clustering method as well as in validating the result. The Gap statistic is one of the most popular techniques to determine the optimal number of clusters. The idea of the Gap statistic is to compare the within-cluster dispersion to its expectation under an appropriate null reference distribution (Tibshirani et al., 2001). It outperforms many other methods, including the method by Kaufman and Rousseeuw (1990), the Caliński and Harabasz (1974) index, the Krzanowski and Lai (1988) method, and the Hartigan (1975) statistic (Tibshirani et al., 2001). Therefore, the Gap statistic is frequently used in a variety of applications, from image segmentation (Zheng-Jun and Yao-Qin, 2009), image edge detection (Yang et al., 2009) to genome clustering (Wendl and Yang, 2004).

However, there are few works investigating the method itself. The tendency of the Gap statistic to overestimate the number of clusters was reported by Dudoit and Fridlyand

†*Address for correspondence:* Mojgan Mohajer, Helmholtz Zentrum München, Ingolstädter Landstr. 1, 85764 Neuherberg, Germany
E-mail: mojgan.mohajer@helmholtz-muenchen.de

(2002). It is also known that the Gap statistic may not work correctly in cases where data are derived from exponential distributions (Sugar and James, 2003). The weighted Gap statistic, proposed by Yan and Ye (2007), is an improvement, for example in the case of mixtures of exponential distribution. Yin et al. (2008) pointed out that in situations where a data set contains clusters of different densities the Gap statistic might fail. They suggested to use reference data sets sampled from normal distribution rather than uniform distribution.

The original Gap statistic is based on some empirical choices, such as the "one standard error" -style rule for simulation error, and using the logarithm of the within cluster dispersion $W_k$. However, few studies have focused on analyzing the effect of these choices. In this paper we will show that using the logarithm of $W_k$ is actually disadvantageous for finding the number of clusters in data sets. Especially in cases where clustering data are sampled from multi-dimensional uniform distributions with large differences in the variances of the different clusters, it is better to use $W_k$ instead of $\log(W_k)$.

The paper is organized as follows. In section 2 the original Gap statistic is described and the difference between the use of the logarithm of $W_k$ and the calculation of the Gap statistic directly from $W_k$ is discussed. In section 3 both Gap functions, with and without log function, are applied to simulated and real data, using hierarchical clustering with average linkage method. We end with a discussion of the results and the proposed method.

## 2. Theory

### 2.1. Gap statistic

Let $\{x_{ij}\}$ be observations with $i = 1, 2, ..., n$, $j = 1, 2, ..., p$, $p$ features measured on $n$ independent samples, clustered into $k$ clusters $C_1, C_2, ..., C_k$, where $C_r$ denotes the indexes of samples in cluster $r$, and $n_r = |C_r|$. Let $d_{ii'}$ be the distance between samples $i$ and $i'$. For example, this distance might be the squared Euclidean distance $d_{ii'} = \sum_j (x_{ij} - x_{i'j})^2$. The sum of the pairwise distances $D_r$ for all points in cluster $r$ is

$$D_r = \sum_{i, i' \in C_r} d_{ii'}. \tag{1}$$

We define

$$W_k := \sum_{r=1}^{k} \frac{1}{2n_r} D_r. \tag{2}$$

If $d$ is the squared Euclidean distance, then $W_k$ is the within-cluster sum of squared distances from the cluster means. $W_k$ decreases monotonically as the number of clusters $k$ increases. For the calculation of the Gap function, Tibshirani et al. (2001) proposed to use the difference of the expected value of $\log(W_k^*)$ of an appropriate null reference and the $\log(W_k)$ of the data set,

$$Gap_n(k) := E_n^* \log(W_k^*) - \log(W_k). \tag{3}$$

Then, the proper number of clusters for the given data set is the smallest $k$ such that

$$Gap_n(k) \geq Gap_n(k+1) - s_{k+1} \tag{4}$$

where $s_k$ is the simulation error calculated from the standard deviation $sd(k)$ of $B$ Monte Carlo replicates $\log(W_k^*)$ according to the equation $s_k = \sqrt{1 + 1/B}\,sd(k)$. The expected value $E_n^* \log(W_k^*)$ of within-dispersion measures $W_{kb}^*$ is determined as

$$E_n^* \log(W_k^*) = \frac{1}{B} \sum_b \log(W_{kb}^*),$$
(5)

where $W_{kb}^*$ are given by clustering the $B$ reference data sets. The sum of $\log(W_{kb}^*)$ can be written as

$$E_n^* \log(W_k^*) = \frac{1}{B} \log \left( \prod W_{kb}^* \right).$$
(6)

Therefore the Gap function from Eqn. 3 can be re-written;

$$Gap_n(k) = \log \left( \frac{(\prod W_{kb}^*)^{1/B}}{W_k} \right).$$
(7)

The number $(\prod W_{kb}^*)^{1/B}$ is the geometric mean of $W_{kb}^*$. Thus, the Gap statistic is the logarithm of the ratio of the geometric mean of $W_{kb}^*$ to $W_k$. In the next section, we will compare this to using the differences of the arithmetic mean of $W_{kb}^*$ and $W_k$.

## 2.2. Gap statistic without logarithm function

Lets considering using $W_k$ instead of $\log(W_k)$. That is, we use an alternative definition of the Gap function,

$$Gap_n^*(k) = E_n^*(W_k^*) - W_k,$$
(8)

where

$$E_n^*(W_k^*) = \frac{1}{B} \sum_b W_{kb}^*.$$
(9)

We refer to the proposed alternative Gap statistic defined by using $W_k$ directly as $Gap_n^*$; the original Gap calculated using the logarithm of $W_k$ is referred to as $Gap_n$. Tibshirani et al. (2001) note that in case of a special Gaussian mixture model $\log(W_k)$ has interpretation as log-likelihood (Scott and Symons, 1971). In maximum likelihood inference, it is usually more convenient to work with the log-likelihood function than with the likelihood function, in order to have sums instead of products. However, using $\log(W_k)$ has no computational advantage versus using $W_k$ directly in the definition of the Gap statistic.

It can be shown that an answer in the original $Gap_n$ is a sufficient condition for the proposed $Gap_n^*$ statistic, but not *vice versa*. Let $A = \prod W_{kb}^{*\,1/B}$, $B = \prod W_{k+1b}^{*\,1/B}$, $C = \frac{1}{B} \sum_b W_{kb}^*$, $D = \frac{1}{B} \sum_b W_{k+1b}^*$, $d_1 = W_k$, and $d_2 = W_{k+1}$.

PROPOSITION 1. *For* $\forall d_1, d_2 > 0$, $d_1 > d_2$, $A > B$, $C > D$, $A, C > d_1$ *and* $B, D > d_2$, *if*

$$log \left( \frac{A}{d_1} \right) \geq log \left( \frac{B}{d_2} \right),$$

*then*

$$C - d_1 \geq D - d_2.$$

PROPOSITION 2. $\exists d_1, d_2 > 0,\ d_1 > d_2,\ A > B,\ C > D,\ A, C > d_1\ and\ B, D > d_2\ so$ *that if*

$$C - d_1 \geq D - d_2,$$

*then*

$$log\left(\frac{A}{d_1}\right) < log\left(\frac{B}{d_2}\right).$$

Proofs are given in Appendix A.

Hence, if there is a possible candidate in $Gap_n$ at point $k$, it is also a possible candidate in $Gap_n^*$. On the other hand it is possible that there is no such $k$ in $Gap_n$ function while the $Gap_n^*$ function indicates a possible candidate at point $k$. In the section 3.4 and 3.5 there are examples from real and simulated data, in which the original $Gap_n$ function is a strictly increasing function, thus there is no $k$ that fulfills the condition in Eqn. 4. However, the proposed $Gap_n^*$ function may be able to suggest a number of clusters for these data sets.

### 2.3. Weighted Gap statistic

In Eqn. 2 $W_k$ is the pooled within-cluster sum of squares. This implies considering a point far away from the cluster mean, the large distance of this point to the cluster center has more impact compared to points with small distances from the cluster mean. To this end, Yan and Ye (2007) suggested to compute $W_k^{'}$ as average of all pairwise distances for all points in a cluster,

$$W_k^{'} = \sum_{r=1}^{k} \frac{2}{n_r(n_r - 1)} D_r. \tag{10}$$

This approach is called "weighted Gap function".

Similar to the original Gap function, the weighted Gap function can also be computed with or without logarithm. However, $W_k$ in Eqn. 2 is monotonically decreasing in $k$ if the distance $d_{ii'}$ is the Euclidean distance. On the other hand, $W_k^{'}$ in Eqn. 10 is not a decreasing (or increasing) function in $k$. Therefore, the propositions given in section 2.2 are not valid for the weighted Gap method. We will compare results from the original and the weighted Gap function on two historical data sets in section 3.1.

## 3. Application to simulated and real data sets

In the previous section we discussed the differences of the Gap functions computed with and without logarithm. In this section we will apply the original *Gap* and proposed *Gap\** statistics to simulated and real data sets, in order to evaluate the effect of the differences in both approaches.

Here, we use agglomerative hierarchical clustering with group average linkage method (Kaufman and Rousseeuw, 1990). The average linkage method has some advantages over the widely used k-mean clustering. Hierarchical clustering methods produce hierarchical representations in which the clusters at each level of the hierarchy are created by merging clusters at the next lower level. Each level of hierarchy represents a particular grouping of the data into disjoint clusters of samples. The entire hierarchy represents an ordered sequence of such groupings. Unlike k-mean clustering, where the choice of different numbers of clusters can lead to totally different assignment of elements to the clusters, in hierarchical clustering the sets of clusters are nested within one another. The average linkage method has another interesting property: the group average dissimilarity $d(G, H)$ between two groups $G$ and $H$ is defined as:

$$d(G, H) = \frac{1}{N_G N_H} \sum_{i \in G} \sum_{i' \in H} d_{ii'}, \tag{11}$$

where $N_G$ and $N_H$ are the number of samples in each group. The group average dissimilarity is an estimate of

$$\int \int d(x, x') p_G(x) p_H(x') dx dx' \tag{12}$$

with the number of observations $N \to \infty$, where $d(x, x')$ is the dissimilarity between points $x$ and $x'$. Eqn. 12 is an approximation for $d(G, H)$, Eqn. 11, when $N$ approaches infinity. This is a characteristic of the relationship between the two densities $p_G(x)$ and $p_H(x')$ of samples in group $G$ and $H$. The average linkage method attempts to produce relatively compact clusters that are relatively far apart (Kaufman and Rousseeuw, 1990).

**Table 1.** Results of standard and weighted $Gap$ and $Gap^*$ functions on Iris and Breast Cancer data sets. "+" indicates the correct number of clusters for that data set.

| Gap function | number of clusters | |
|---|---|---|
| | Iris | Breast |
| $Gap$ | $3^+$ | $2^+$ |
| $Gap^*$ | $3^+$ | $2^+$ |
| $weighted\_Gap$ | 2 | 1 |
| $weighted\_Gap^*$ | 7 | 1 |

### 3.1. Two historical data sets

Two historical data sets are frequently used when discussing clustering; "Fisher's Iris data set" (Fisher, 1963) and Wolbergs "Breast Cancer Wisconsin data set" (Wolberg et al., 1993). We apply the four different definitions of the Gap statistic to these two famous historical data sets. Fisher's Iris data set consists of 50 samples from three species of Iris flowers. Four variables were measured for each sample. For the "Breast Cancer Wisconsin data set", samples arrived periodically as Dr. Wolberg reports his clinical cases. The data set consists of 699 samples. Each sample is described by nine variables. The whole data set has two main groups, consisting of 458 benign and 241 malignant tumors.

Table 1 lists the estimated number of clusters for both the iris and the breast data sets using the original Gap statistic $Gap$ from Eqn. 3 and the proposed Gap statistic without logarithm $Gap^*$ as defined in Eqn. 8. These two Gap functions are compared with the results of the $weighted\_Gap$ as described in section 2.3 and the $weighted\_Gap^*$, *i.e.*, the weighted Gap using $W_k$ instead of $\log(W_k)$.

In contrast to the result from k-mean clustering reported by Yan and Ye (2007), when using average linkage clustering the Gap statistic with the original $W_k$, Eqn. 2, estimates the number of clusters for both data sets correctly. Figs. 1 and 2 show the calculated Gap functions for the two data sets. Both, the iris and the breast cancer data sets represent their natural clusters in average linkage clustering. Thus, $Gap$ and $Gap^*$ show similar behavior. It can be observed that in the case of iris data, the $weighted\_Gap$ suggests number 2 as proper number of clusters but $weighted\_Gap^*$ suggest 7 as cluster number. According to the discussion in section 2.2, whenever a number fulfills the inequality 4, this number fulfills the inequality for the proposed $Gap^*$. However, this statement is not valid for $weighted\_Gap$ due to the fact that $W_k'$ from Eqn. 10 is not monotonically decreasing.

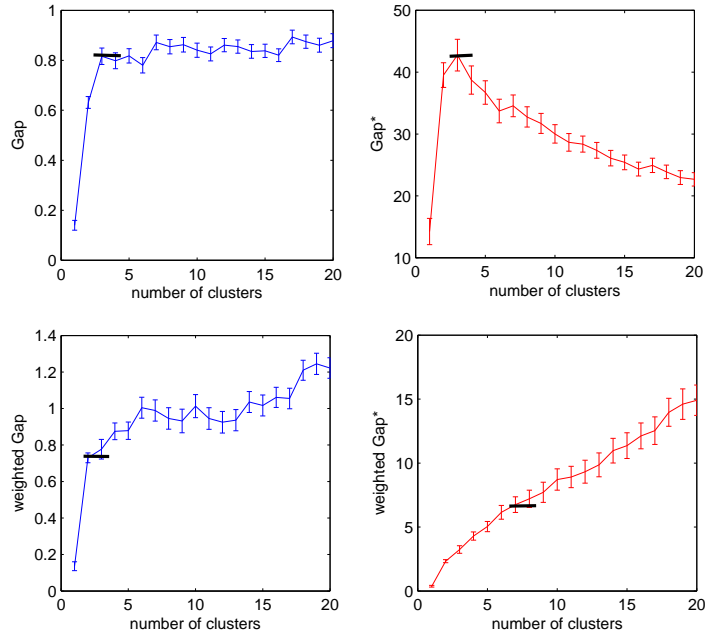

**Fig. 1.** Standard and weighted $Gap$ and $Gap^*$ functions for Iris data set

### 3.2. Not well separated clusters

Now we assume clusters which are not well separated. We simulated 1000 data sets with two clusters each, with different proportions of overlapping. Each cluster had 50 observations with two variables. Both variables were drawn independently from Gaussian distributions; for observations from the first cluster both variables had expected values 0 and standard deviation 1. For observations from the second cluster both variables were again randomly drawn from Gaussian distribution with expected value $\Delta$ and standard deviation 1. As a result, there are two clusters, where the distance between the means of two clusters decreases with decreasing value of $\Delta$. We use $\Delta = 0.5, 1, 1.5, \ldots, 5.0$. For each of the ten unique values of $\Delta$ 100 data sets were generated, and original $Gap$ and proposed $Gap^*$ functions were calculated for these data sets. Figure 3 shows the percentage of finding two as the number of clusters for each type of data set. It can be observed that the original $Gap$ was better in estimating the proper number of clusters in overlapped clusters than $Gap^*$. These results were expected due to the tendency of the $Gap$ to overestimate the number of clusters which has been reported by Dudoit and Fridlyand (2002).

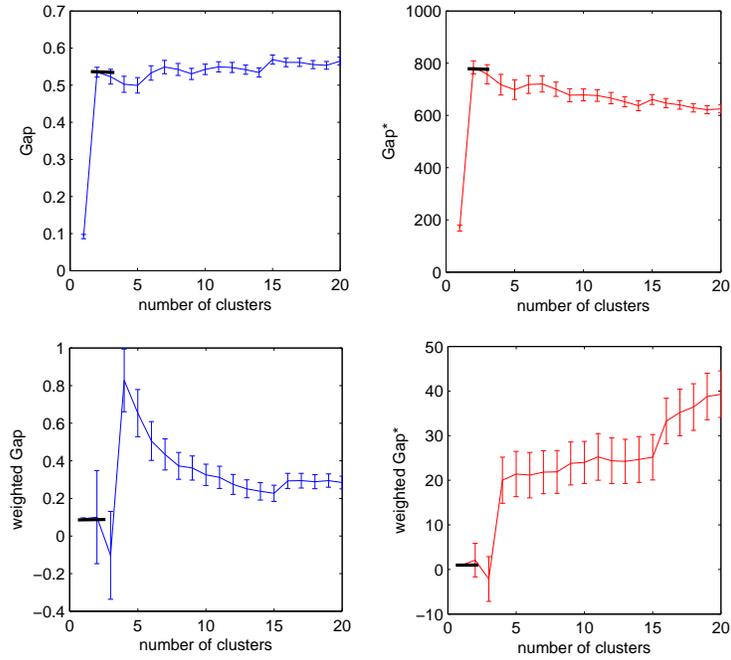

**Fig. 2.** Standard and weighted $Gap$ and $Gap^*$ functions for Breast Cancer Wisconsin data set

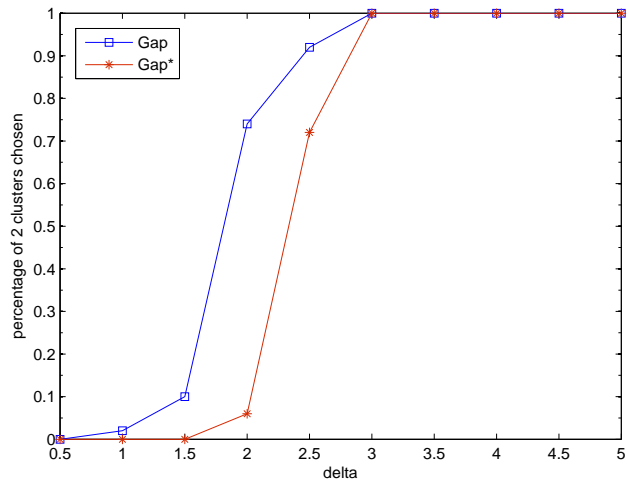

**Fig. 3.** $Gap$ function from eq. 3 and $Gap^*$ function from eq. 8 are compared for 10 data sets with two clusters. Two clusters have different portion of overlapping area in each data set.

**Table 2.** Five simulated data sets with two clusters with $N_1$ and $N_2$ number of samples in first and second cluster respectively.

| simulation | $N_1$ | $N_2$ | $m = N_1/N_2$ |
|------------|-------|-------|---------------|
| 1 | 765 | 765 | 1 |
| 2 | 1020 | 510 | 2 |
| 3 | 1224 | 306 | 4 |
| 4 | 1360 | 170 | 8 |
| 5 | 1440 | 90 | 16 |

### 3.3. Unequally sized clusters

Yin et al. (2008) report that whenever the number of observations in one cluster is more than six-fold the number of observations in the other clusters, the Gap statistic is not able to estimate the number of clusters accurately. This effect depends not only on the number difference between clusters but also on the distance between clusters. We study this effect in the special case of two clusters sampled from two 2D normal distributions $N(\boldsymbol{\mu}, \boldsymbol{I})$ and $N(\boldsymbol{\mu}', \boldsymbol{I})$, where $\boldsymbol{\mu}$ and $\boldsymbol{\mu}'$ are two different expected values and $\boldsymbol{I}$ is the identity matrix. Details of this study are given in Appendix B. Suppose $N_1$ is the number of samples in the first cluster and $N_2$ is the number of samples in the second cluster and $N_1 = m \cdot N_2$ and $n = N_1 + N_2$. For a fixed total number of samples $n$, by increasing $m$, the value of $W_1$ decreases. Thus, $Gap_1$ increases while $Gap_2$ is almost unchanged. When $m$ becomes large enough, $Gap_1$ will be greater than $Gap_2$, and the estimated cluster number will be one. The possible numbers of $m$ for which $Gap$ and $Gap^*$ can still estimate two as proper number of clusters, can be estimated from the following two inequalities (see Appendix B inequalities Eqns. 24 and 25):

(a) for $Gap$

$$\frac{md}{(m+1)^2} \geq \frac{E(d_1)}{E(d_2)} - 1 \tag{13}$$

(b) for $Gap^*$

$$\frac{2md}{(m+1)^2} \geq E(d_1) - E(d_2) \tag{14}$$

where $d$ is the average distance between the points in first cluster to the points in second cluster, $E(d_1)$ is the expected distance of two points from a rectangular uniform distribution with sides $a$ and $b$ and $E(d_2)$ is the expected distance of two points from a rectangular uniform distribution with sides $\frac{a}{2}$ and $b$.

These results are illustrated in an example in Fig. 4. In this example we compared five data sets with two clusters of different observation sizes. The total number of observations is the same in all five data sets, however, the ratio of observations is varied. Table 2 summarizes the size of the clusters in each data set and the ratio between number of observations in the two clusters. In the first data set the number of observations in the first ($N_1$) and in the second cluster ($N2$) are equal. In the other four data sets $N_1$ increases and $N_2$ decreases as given in table 2.

Samples where drawn as follows:

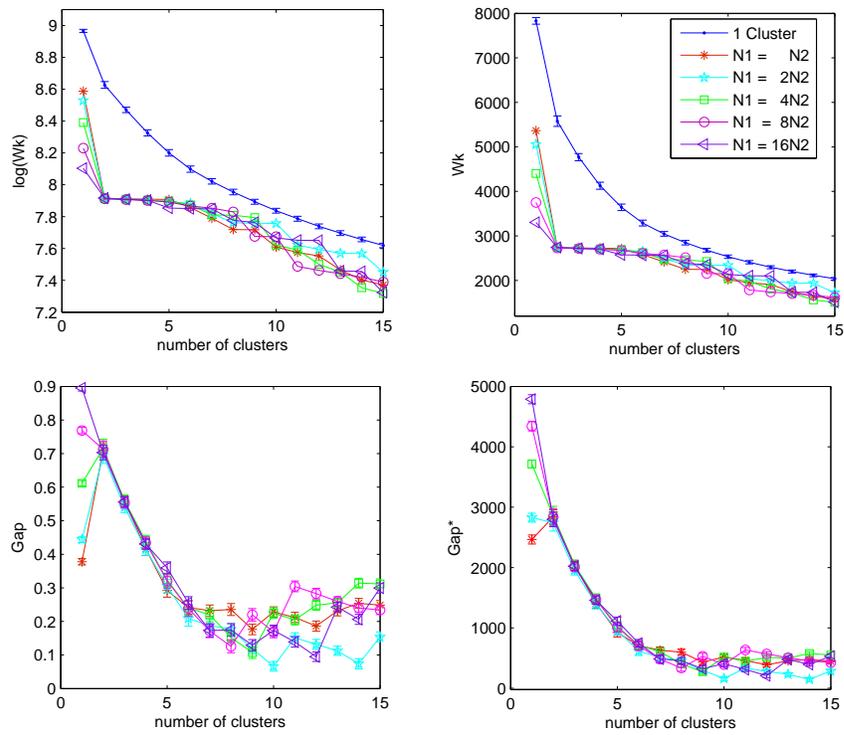

**Fig. 4.** Top: $\log(W_k)$ (left) and $W_k$ (right) five simulated data sets with two clusters each, where $N1$ and $N2$ are the number of samples in the first and second cluster, respectively. Bottom: $Gap$ (left) and $Gap^*$ (right) for these data sets.

(a) Select $N_1^{max}$ as maximum number of samples in first cluster in all five data sets.

(b) Select $N_2^{max}$ as maximum number of samples in second cluster in all five data sets.

(c) Draw $N_1^{max}$ samples from a bivariate normal distribution with parameters $(\boldsymbol{\mu}, \boldsymbol{I})$, where $\boldsymbol{\mu} = (0, 0)$.

(d) Draw $N_2^{max}$ samples from a bivariate normal distribution with parameters $(\boldsymbol{\mu'}, \boldsymbol{I})$, where $\boldsymbol{\mu'} = (5, 0)$.

(e) For each data set, select the first $N_1$ samples from the $N_1^{max}$ sample points according to the number $N_1$ given for this data set in table 2.

(f) For each data set, select the first $N_2$ samples from the $N_2^{max}$ sample points according to the number $N_2$ given for this data set in table 2.

According to the estimations in Appendix B and the inequalities 13 and 14, in this example, $E(d_1) \approx 4.53$, $E(d_2) \approx 2.99$, and $d \approx 3.48$. As a result only for $m < 6$ for the original *Gap*, and $m < 2$ for the proposed *Gap**, the gap statistic determines two as proper number of clusters. Figure 4 shows $\log(W_k)$ and $W_k$ for all five simulated data sets. The blue dotted line is the expected $\log(W_k)$ on the left top and expected $W_k$ on the right top of the null reference distribution. As demonstrated in figure 4, by increasing the number of samples in first cluster against the second cluster, the within-cluster dispersion $W_2$ remains the same but $W_1$ decreases. Depending on how far apart the two clusters are, increasing the ratio of observations in both clusters increases the $Gap(1)$ value. Figure 4 demonstrates the original *Gap* function (bottom left) and the proposed *Gap** function (bottom right) for these five data sets. The estimated $m$ from the inequalities 13 and 14 is confirmed by the results illustrated in Fig. 4.

### 3.4.  *Simulated data with increasing Gap function*

In this experiment, data were simulated such that the calculated Gap function (*Gap* from Eqn. 3) is a strictly increasing function. A data set was simulated 2000 times and for each simulated data set the original *Gap* and the proposed *Gap** statistic was calculated. The simulated data set consists of two clusters each. Each cluster contains 50 observations from an n-dimensional variable space. In the first cluster, each feature was sampled from a uniform distribution on interval $[0, 10]$ at random. For the second cluster only the first variable was sampled from the same uniform distribution. All other variables of observations in the second cluster were set to zero. Half of the data sets were simulated in a 100-dimensional variable space while the other half were simulated in a 2-dimensional variable space.

Figure 5 depicts the average *Gap* and the average *Gap** functions for both the 2D data sets and the 100D data sets. For the 2D data sets, both Gap functions suggest two as proper number of clusters. However, it can be seen that the *Gap* function for the 100D data sets is a strictly increasing function. This is indeed expected due to the "curse of dimensionality" (Bellman, 1961). Beyer et al. (1999) have shown that the minimum and the maximum occurring distances become indiscernible, as the difference of the minimum and maximum value compared to the minimum value converges to 0 as the dimensionality $d$ goes to infinity.

$$\lim_{d \to \infty} \frac{dist_{max} - dist_{min}}{dist_{min}} \to 0. \tag{15}$$

Consequently, all the distances $d_{ii'}$ from Eqn. 1 can be considered to be equal in a high dimensional space. Consider $n$ observations from a 100 dimensional uniform distribution and suppose these samples are divided into $k$ clusters $C_1, C_2, ..., C_k$, where $|C_1| = |C_2| = ... = |C_k| = \frac{n}{k}$. Consider all $d_{ii'} = dist$, thus, $W_k$ is equal to:

$$W_k = \left( \frac{n}{2} - \frac{k}{2} \right) dist. \tag{16}$$

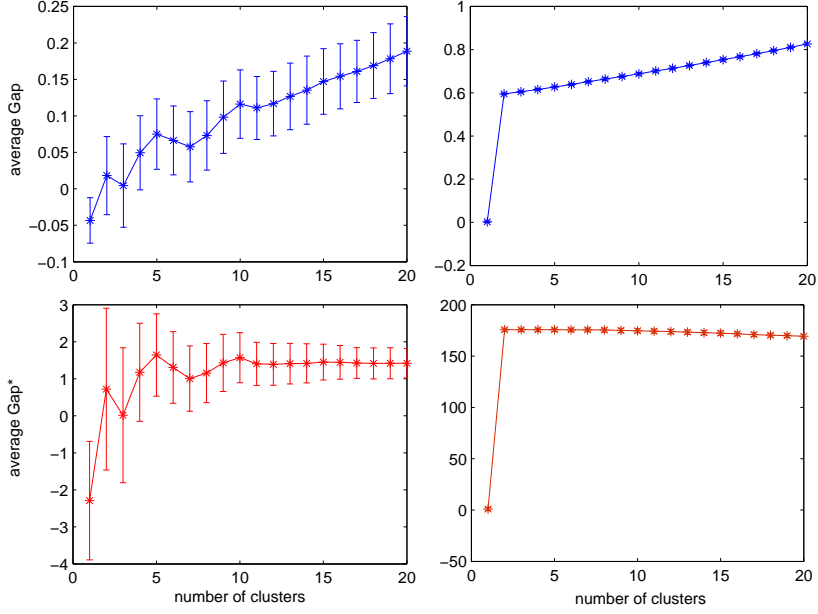

**Fig. 5.** Average $Gap$ and $Gap^*$ for simulated 2D (left) and 100D (right) data sets from experiments 3.4.

By increasing the number of clusters $k$, $W_k$ in Eqn. 16 decreases linearly. The slope of this line is the same for all data sets sampled from the same high dimensional uniform population even with different number of samples. Here, in the case of a 100D data set for all $k > 2$ only the first cluster will be divided further, due to the large distances of the samples in this cluster compared to the second cluster. Hence, $W_k$ will be linear for $k > 2$ and parallel to $E^*(W_k^*)$. The difference $E^*(W_k^*) - W_k$ remains constant as $E^*(W_k^*)$ and $W_k$ decrease. Therefore, the $Gap$ function is strictly increasing. On the other hand, whenever the difference $E^*(W_k^*) - W_k$ remains constant, $Gap^*(k)$ and $Gap^*(k+1)$ will be equal. Therefore, due to the Gap condition Eqn. 4 $k$ will be suggested as proper number of clusters by the proposed $Gap^*$ statistic.

Table 3 lists the number of clusters found with the original $Gap$ and the proposed $Gap^*$ statistic for 1000 simulations of 2D and 100D data sets, respectively. While for the 2D simulation both the original $Gap$ and the proposed $Gap^*$ statistic perform similarly, the original $Gap$ fails in finding the true number of clusters for all of the 1000 simulated 100D data sets. The proposed $Gap^*$ statistic, however, is able to determine the true number of clusters for these simulations.

**Table 3.** Number of clusters for 1000 2D and 100D data sets, estimated by $Gap$ and $Gap^*$.

| | Method | \multicolumn{11}{c}{Estimate of number of clusters} |
|---|---|---|---|---|---|---|---|---|---|---|---|
| | | 1 | 2 | 3 | 4 | 5 | 6 | 7 | 8 | 9 | $\geq 10$ |
| 2D | $Gap$ | 368 | 489 | 143 | 0 | 0 | 0 | 0 | 0 | 0 | 0 |
| | $Gap^*$ | 270 | 567 | 162 | 1 | 0 | 0 | 0 | 0 | 0 | 0 |
| 100D | $Gap$ | 0 | 0 | 0 | 0 | 0 | 0 | 1 | 3 | 1 | 995 |
| | $Gap^*$ | 0 | 1000 | 0 | 0 | 0 | 0 | 0 | 0 | 0 | 0 |

### 3.5. Real data set with increasing Gap function

We evaluated both Gap functions further on seven real data sets from Dynamic Contrast-Enhanced Magnetic Resonance Imaging (DCE-MRI) of breast tumors (German Cancer Research Center (DKFZ), 2004). For each data set a selected slice through the tumor with thickness $TH = 6mm$ and field of view $FOV = 320mm \times 320mm$ was measured every $3.25s$ for 6.9 minutes. As a result, each voxel in a data set is described by a signal time curve of length $T = 128$ during the contrast agent passage through the tumor (Brix et al., 2004). These curves give valuable information about blood circulation and permeability of tumor tissue. Hence, it is of interest to detect voxels with similar signal curves. Previously different clustering methods were applied on DCE-MRI data (Fischer and Hennig, 1999; Nattkemper et al., 2005; Varini et al., 2006; Wismüller et al., 2006; Schlossbauer et al., 2008; Castellani et al., 2009). One of the main challenges on this approach is to determine the number of underlying patterns in the signal curves. To this end we applied the Gap statistic on DCE-MRI data. As before, we used the average linkage clustering method with squared Euclidean distance as measure of dissimilarity. The samples are the signal curves of voxels of which each is described by 128 features, *i.e.*, time points.

Table 4 gives the number of clusters found with the original $Gap$ and the proposed $Gap^*$ for seven DCE-MRI data sets. The tumors in all of these images have the same type. Using the proposed $Gap^*$ statistic, the number of five clusters was found in five of the seven images, whereas with the original $Gap$ statistic, no consistent number of clusters, *i.e.*, regions, was found.

Fig. 6 shows the resulting $Gap$ and $Gap^*$ functions for one of the DCE-MR images (data set 4). Similar to the simulated data set in 3.4, the $Gap$ function is a strictly increasing function, whereas the $Gap^*$ function is not strictly increasing and suggests five as number of clusters for this data set. In Fig. 7(a) first and second principal component of the data set are depicted and the five identified clusters are shown in different colors and with different symbols. The intensity curves for voxels in a cluster are shown in Fig. 7(c); the mean curve of each cluster is depicted in red. Fig. 7(b) depicts the tumor image with voxel colored according to their cluster with the same colors as in sub-figure (a). A ring-shaped ordering of the five clusters can be observed in this image. This ordering is in agreement with enhancement patterns reported in medicine such as, circumferential, centripetal and peripheral ring contrast (Buadu et al., 1997). However, so far there is no information on the number of regions.

## 4. Discussion

The Gap statistic is one of the most popular methods for estimating the number of clusters in a data set. It is rather simple to implement and is used in many, diverse applications.

**Table 4.** Results for all seven DCI-MRI data sets analyzed with the $Gap$ and the $Gap^*$ statistic. $nd$ stands for not defined.

| data set | number of voxels | $Gap$ | $Gap^*$ |
|----------|-----------------|-------|---------|
| 1 | 1260 | 7 | 7 |
| 2 | 207 | 9 | 5 |
| 3 | 116 | 9 | 5 |
| 4 | 262 | $nd$ | 5 |
| 5 | 141 | 11 | 5 |
| 6 | 277 | $nd$ | 5 |
| 7 | 151 | 13 | 4 |

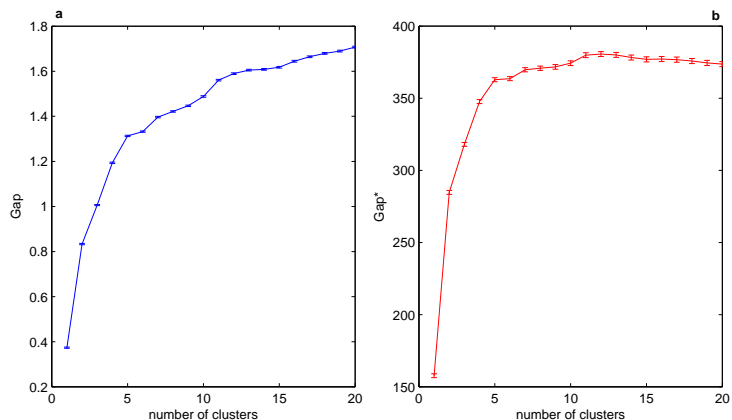

**Fig. 6.** Gap functions $Gap$ and $Gap^*$ for DCE-MRI data set of a breast tumor.

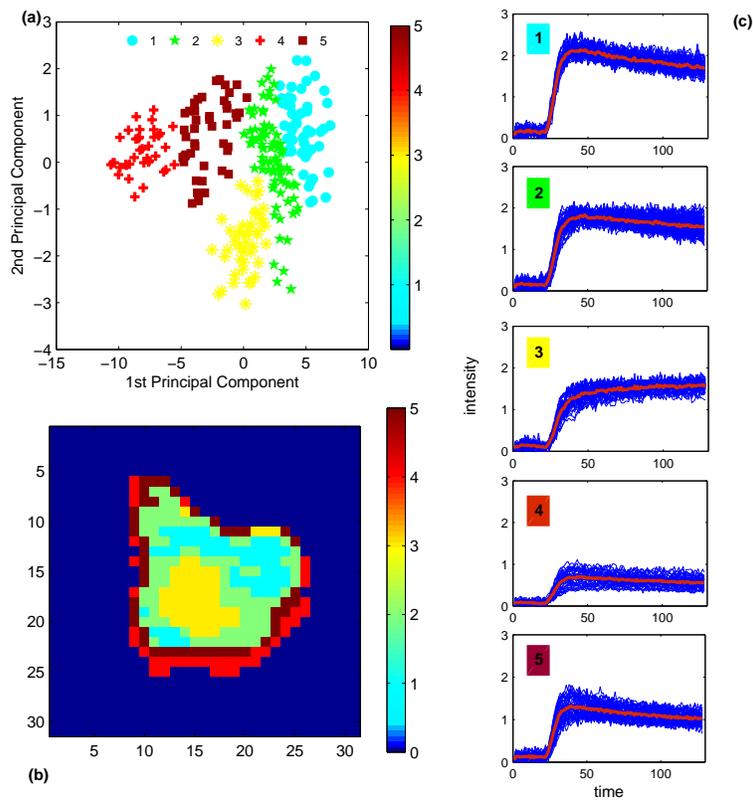

**Fig. 7.** Five clusters of the DCE-MRI breast tumor found with average linkage clustering. (a) First and second principal component of the DCE-MRI signals per voxel. Voxels are colored according to their cluster affiliation. (b) Segmentation map of the tumor. Voxels are colored similar to subfigure (a). (c) Signal time curves for each voxel in the five respective clusters along with the mean curve (bold red line).

As reported by Tibshirani et al. (2001) it outperforms many other methods. However the Gap statistic is not able to suggest the correct number of clusters in some cases. Yin et al. (2008) have reported that in cases where the ratio of observation sizes between clusters is over than six-fold, the Gap statistic does not work accurately. Dudoit and Fridlyand (2002) have mentioned the overestimation of Gap statistic in some applications. Sugar and James (2003) have reported the failure of the Gap statistic in the case that data were derived from exponential distributions.

In this paper we have shown that using $\log(W_k)$ instead of $W_k$ in the calculation of the Gap function can be one cause of overestimation of number of clusters in the Gap statistic. Theoretically there is no feasible reason to choose Eqn. 3 over Eqn. 8 for the definition of the Gap statistic. Indeed, using the logarithm function in the definition of the *Gap* statistic has a fundamental effect on the results of the *Gap* statistic. This is due to a property of the logarithm function described in following example: Consider four positive numbers $a$, $b$, $c$, and $d$, with logarithm of all of them greater than 1. Let be $a > c$ and $b > d$ and $a - b = c - d > 0$, then we will have $\log(a) - \log(b) < \log(c) - \log(d)$. As a result, by increasing the number of clusters the within cluster dispersion $W_k$ decreases. Consequently the *Gap* function increases even when the distance between $W_k^*$ and $W_k$ remains the same.

Estimating the number of clusters depends on many factors. The choice of clustering method is one of these factors. The Gap statistic is designed to be applicable to any clustering method. In general, the results and discussions given in this work are not restricted to any clustering method. However, the choice of the clustering method influences the result of Gap statistic. Different clustering methods look for different structures in data. The average linkage method, used for the Gap calculation in section 3.1, was able to find the real cluster number for both the "iris" (Fisher, 1963) and the "Breast Cancer Wisconsin data set" (Wolberg et al., 1993) in contrast to the Gap function with k-mean clustering reported by Yan and Ye (2007).

Comparing the original *Gap* and proposed *Gap** statistic, the original *Gap* statistic has a better performance in the case of overlapped clusters than *Gap** due to the tendency of the *Gap* of overestimating the number of clusters. For real application, it is however up to the user to decide whether two clusters with overlapping area should be considered as one cluster or two. In previous studies (Tibshirani et al., 2001; Yan and Ye, 2007; Yin et al., 2008; Dudoit and Fridlyand, 2002; Sugar and James, 2003) it was reported that a null reference data generated from a uniform distribution aligned with the principal components of the data causes a better performance of *Gap* statistic. The *Gap* function calculated from such null reference data is referred to as $Gap_{pc}$. It would be interesting to compare $Gap_{pc}$ and $Gap_{pc}^*$ in further studies.

We have introduced *Gap**, which compares the expected values of $W_k^*$ with $W_k$. Thus, it reflects exactly the changes in the within cluster dispersion of the real data against the expected $W_k^*$ of the null reference data set. Whenever the original *Gap* results in a $k$ as proper number of cluster, this $k$ is also a possible answer with the proposed *Gap**. In contrast, there are situations where proposed *Gap** function is able to offer a number as a proper number of clusters while the original *Gap* has no answer. Evaluations in section 3 verify this idea. In subsections 3.4 and 3.5, the original *Gap* function is a strictly increasing function, hence it cannot find any cluster number. On the other hand, *Gap** is not strictly increasing and therefore is able to suggest a cluster number for the data. For the simulated data in subsection 3.4 the suggested number is equal to real number of clusters. For the real data set in subsection 3.5, however, we have no reference to decide if the number suggested by the proposed *Gap** statistic is the proper number of clusters. Further experiments are necessary on real data with known cluster number to verify the accuracy of the proposed *Gap\** statistic in cases where the original *Gap* is a strictly increasing function. Our experiments suggest that such data are possibly from multi dimensional feature space, with different variances in the different feature axes.

## 5.  Acknowledgments

We are thankful to the UCI Machine Learning Repository which provides a free access to real data sets (`http://archive.ics.uci.edu/ml`). We also thank German Cancer Research Center DKFZ, which provides us with DCE-MRI data sets. This study was performed as part of a joint research project supported by the German "Competence Alliance on Radiation Research." Volker J. Schmid was funded by the LMUinnovative project "BioMed-S – Analysis and Modeling of Complex Systems".

## A.  Proofs

### A.1.  Proof of proposition (1):

PROOF.

$$log(\frac{A}{d_1}) \geq log(\frac{B}{d_2}) \Rightarrow \frac{A}{B} \geq \frac{d_1}{d_2}$$

$\Rightarrow$

$$\frac{A}{B} \geq 1 \text{ and } \frac{d_1}{d_2} \geq 1 \Rightarrow \frac{A}{B} - 1 \geq \frac{d_1}{d_2} - 1$$

Proof by contradiction: If $C - d_1 \geq D - d_2$ is not true, then we have:

$$C - d_1 < D - d_2$$

$$C - D < d_1 - d_2 \Rightarrow \frac{C - D}{d_2} < \frac{d_1}{d_2} - 1$$

$\Rightarrow$

$$\frac{C - D}{d_2} < \frac{A}{B} - 1$$

$$\frac{1}{B}\sum_b W^*_{kb} - \frac{1}{B}\sum_b W^*_{k+1b} < d_2\frac{\prod W^{*\,1/B}_{kb}}{\prod W^{*\,1/B}_{k+1b}} - d_2 \tag{17}$$

Geometric to arithmetic mean relationship says:

$$\left(\prod \frac{W^*_{kb}}{W^*_{k+1b}}\right)^{1/B} \leq \frac{1}{B}\sum_b \frac{W^*_{kb}}{W^*_{k+1b}}$$

so we can rewrite the Eqn. 17 as follows:

$$\frac{1}{d_2}\sum_b W^*_{kb} - \frac{1}{d_2}\sum_b W^*_{k+1b} < \sum_b \frac{W^*_{kb}}{W^*_{k+1b}} - B \tag{18}$$

$\Rightarrow$

$$\sum_b \frac{W^*_{kb}}{d_2} - \sum_b \frac{W^*_{kb}}{W^*_{k+1b}} < \sum_b \frac{W^*_{k+1b}}{d_2} - B$$

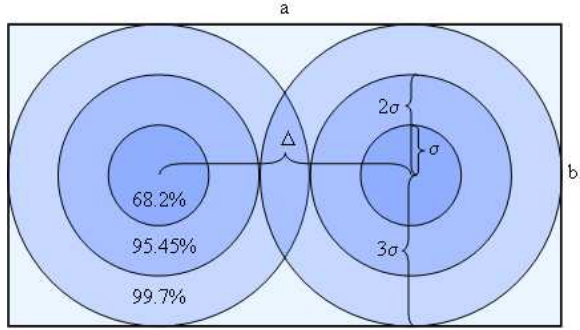

**Fig. 8.** Two 2D distributions from the case study in Appendix B. Each distribution is depicted with three areas of $68.2\%$, $95.45\%$, and $99.7\%$ percentage of sample occurrence inside each area.

$$\Rightarrow$$

$$\sum_b \frac{W_{kb}^*}{W_{k+1b}^*}\left(\frac{W_{k+1b}^* - d_2}{d_2}\right) < \sum_b \frac{W_{k+1b}^* - d_2}{d_2}$$

For all values of $b$, $\frac{W_{kb}^*}{W_{k+1b}^*} \geq 1$, setting $\frac{W_{kb}^*}{W_{k+1b}^*}$ to its minimum value 1, then we have:

$$\sum_b \frac{W_{k+1b}^* - d_2}{d_2} < \sum_b \frac{W_{k+1b}^* - d_2}{d_2}$$

$\square$

### A.2. Proof of proposition (2):

PROOF. From the previous proof we have:

$$\frac{A}{B} - 1 < \frac{C - D}{d_2}$$

Thus, in the case of $d_1 - d_2 = C - D$ we will have:

$$\frac{A}{B} < \frac{d_1}{d_2}$$

$\square$

## B.  Case Study: Unequally sized clusters

In the following case study the effect of number difference between clusters on Gap statistic was studied. The case study considered data sets with each consisting of two clusters sampled from two 2D normal distributions $N(\boldsymbol{\mu}, \sigma^2\mathbf{I})$ and $N(\boldsymbol{\mu}', \sigma^2\mathbf{I})$, where $\boldsymbol{\mu}$ and $\boldsymbol{\mu}'$ are expected values, $\mathbf{I}$ is the identity matrix, and $\sigma^2 > 0$ is a positive real number. According to standard score (Glenberg and Andrzejewski, 2008) $99.7\%$ of samples will be inside a circle with radius $3 \cdot \sigma$. Here, the uniform distribution rectangle, from which the null references are sampled, was estimated as a rectangle with sides $6 \cdot \sigma + \Delta$ and $6 \cdot \sigma$ as illustrated in Fig. 8, where $\Delta = \|\boldsymbol{\mu} - \boldsymbol{\mu}'\|$. Let $N_1$ be the number of samples in the first cluster and $N_2$ be the number of samples in the second cluster, while $N_1 = m \cdot N_2$ and $n = N_1 + N_2$. In section 3.2 we observed that in the case of $N_1 = N_2$, for $\Delta \geq 5\sigma$ both Gap functions estimate two as proper number of clusters. In this study we want to show how changes in $m$ affect the result of the Gap statistic. Let $\Delta \geq 5\sigma$ and $n$ be fixed. For the Gap statistic it is necessary to have $Gap_n(1) \leq Gap_n(2) - s_2$ in order to be able to choose $k = 2$ as proper number of clusters otherwise it suggests $k = 1$. We ignore $s_2$ and consider the inequality $Gap_n(1) \leq Gap_n(2)$. The two next inequalities follow from the Eqns. 3 to 8 for $Gap$ and $Gap^*$, respectively:

(a) *Gap*

$$\left( \prod \frac{W_{1b}^*}{W_{2b}^*} \right)^{\frac{1}{B}} \leq \frac{W_1}{W_2} \tag{19}$$

(b) *Gap*$^*$

$$\frac{1}{B} \sum \left( W_{1b}^* - W_{2b}^* \right) \leq W_1 - W_2 \tag{20}$$

Each $W_{1b}^*$ can be estimated as $nE(d_1)$ where $E(d_1)$ is the expected distance between two random points from a rectangular uniform distribution with sides $6\sigma + \Delta$ and $6\sigma$. In a similar way $W_{2b}^*$ can be estimated as $nE(d_2)$ where $E(d_2)$ is the expected distance between two random points from a rectangular uniform distribution with sides $\frac{6\sigma + \Delta}{2}$ and $6\sigma$. The expected distance of two random points sampled from a rectangular uniform distribution with sides $a$ and $b$ with $a \geq b$ is given by (Santalo, 1976)

$$E(d) = \frac{1}{15} \left[ \frac{a^3}{b^2} + \frac{b^3}{a^2} + d \left( 3 - \frac{a^2}{b^2} - \frac{b^2}{a^2} \right) + \frac{5}{2} \left( \frac{b^2}{a} \log \frac{a + d}{b} + \frac{a^2}{b} log \frac{b + d}{a} \right) \right] \tag{21}$$

where $d = \sqrt{a^2 + b^2}$. Using these estimations and Eqns. 19 and 20 we gain

(a) *Gap*

$$\frac{E(d_1)}{E(d_2)} \leq \frac{W_1}{W_2} \tag{22}$$

(b) *Gap*$^*$

$$n \left( E(d_1) - E(d_2) \right) \leq W_1 - W_2 \tag{23}$$

Furthermore, we can take into account that $W_1$ includes the inter-cluster distances between the first and second clusters in addition to all distances which are used in calculation of $W_2$. Therefore $W_1$ can be written as $W_2 + \frac{2N_1 N_2 d_\Delta}{n}$, where $d_\Delta$ is the average inter-cluster distances. Consequently, inequalities (22) and (23) can be rewritten as:

(a) *Gap*

$$\frac{E(d_1)}{E(d_2)} - 1 \leq \frac{m d_\Delta}{\sigma (m + 1)^2} \tag{24}$$

(b) *Gap*$^*$

$$E(d_1) - E(d_2) \leq \frac{2m d_\Delta}{(m + 1)^2} \tag{25}$$

## References

Bellman, R. (1961). *Adaptive control processes: a guided tour.* A Rand Corporation Research Study Series. Princeton University Press.

Beyer, K., J. Goldstein, R. Ramakrishnan, and U. Shaft (1999). When is nearest neighbor meaningful? In C. Beeri and P. Buneman (Eds.), *Database Theory ICDT99*, Volume 1540 of *Lecture Notes in Computer Science*, pp. 217–235. Springer Berlin / Heidelberg.

Brix, G., F. Kiessling, R. Lucht, S. Darai, K. Wasser, S. Delorme, and J. Griebel (2004). Microcirculation and microvasculature in breast tumors: pharmacokinetic analysis of dynamic MR image series. *Magnetic Resonance in Medicine 52*(2), 420–429.

Buadu, L., J. Murakami, S. Murayama, N. Hashiguchi, S. Sakai, S. Toyoshima, K. Masuda, S. Kuroki, and S. Ohno (1997). Patterns of peripheral enhancement in breast masses: correlation of findings on contrast medium enhanced MRI with histologic features and tumor angiogenesis. *Journal of computer assisted tomography 21*(3), 421.

Caliński, T. and J. Harabasz (1974). A dendrite method for cluster analysis. *Communications in Statistics-Theory and Methods 3*(1), 1–27.

Castellani, U., M. Cristiani, A. Daducci, P. Farace, P. Marzola, V. Murino, and A. Sbarbati (2009). DCE-MRI data analysis for cancer area classification. *Methods of information in medicine 48*(3), 248–253.

Dudoit, S. and J. Fridlyand (2002). A prediction-based resampling method for estimating the number of clusters in a dataset. *Genome biology 3*(7).

Fischer, H. and J. Hennig (1999). Neural network-based analysis of MR time series. *Magnetic Resonance in Medicine 41*(1), 124–131.

Fisher, R. (1963). Irvine, CA: University of California, School of Information and Computer Science: UCI Machine Learning Repository. http://archive.ics.uci.edu/ml.

German Cancer Research Center (DKFZ) (2004). *Research program "Innovative Diagnosis and Therapy".* Heidelberg, Germany: German Cancer Research Center (DKFZ).

Glenberg, A. and M. Andrzejewski (2008). *Learning from data: An introduction to statistical reasoning.* Taylor & Francis Group, LLC.

Hartigan, J. (1975). *Clustering algorithms.* John Wiley & Sons, Inc. New York, NY, USA.

Kaufman, L. and P. Rousseeuw (1990). *Finding Groups in Data An Introduction to Cluster Analysis.* New York: Wiley Interscience.

Krzanowski, W. and Y. Lai (1988). A criterion for determining the number of groups in a data set using sum-of-squares clustering. *Biometrics 44*(1), 23–34.

Nattkemper, T., B. Arnrich, O. Lichte, W. Timm, A. Degenhard, L. Pointon, C. Hayes, and M. Leach (2005). Evaluation of radiological features for breast tumour classification in clinical screening with machine learning methods. *Artificial Intelligence in Medicine 34*(2), 129–139.

Santalo, L. A. (1976). *Integral geometry and geometric probability / Luis A. Santalo ; with a foreword by Mark Kac*, pp. 49. Addison-Wesley Pub. Co., Advanced Book Program, Reading, Mass. :.

Schlossbauer, T., G. Leinsinger, A. Wismuller, O. Lange, M. Scherr, A. Meyer-Baese, and M. Reiser (2008). Classification of small contrast enhancing breast lesions in dynamic magnetic resonance imaging using a combination of morphological criteria and dynamic analysis based on unsupervised vector-quantization. *Investigative radiology 43*(1), 56.

Scott, A. and M. Symons (1971). Clustering methods based on likelihood ratio criteria. *Biometrics 27*(2), 387–397.

Sugar, C. A. and G. M. James (2003). Finding the Number of Clusters in a Data Set - An Information Theoretic Approach. *J. Am. Statist. Ass. 98*(463), 750–763.

Tibshirani, R., G. Walther, and T. Hastie (2001). Estimating the number of clusters in a data set via the gap statistic. *J. R. Statist. Soc. B 63*(2), 411–423.

Varini, C., A. Degenhard, and T. Nattkemper (2006). Visual exploratory analysis of DCE-MRI data in breast cancer by dimensional data reduction: A comparative study. *Biomedical Signal Processing and Control 1*(1), 56–63.

Wendl, M. and S. Yang (2004). Gap statistics for whole genome shotgun DNA sequencing projects. *Bioinformatics 20*(10), 1527–1534.

Wismüller, A., A. Meyer-Bäse, O. Lange, T. Schlossbauer, M. Kallergi, M. Reiser, and G. Leinsinger (2006). Segmentation and classification of dynamic breast magnetic resonance image data. *Journal of Electronic Imaging 15*, 013020.

Wolberg, W., W. Street, and O. Mangasarian (1993). Irvine, CA: University of California, School of Information and Computer Science: UCI Machine Learning Repository. http://archive.ics.uci.edu/ml.

Yan, M. and K. Ye (2007). Determining the number of clusters using the weighted gap statistic. *Biometrics 63*(4), 1031–1037.

Yang, Q., L. Tang, W. Dong, and Y. Sun (2009). Image edge detecting based on gap statistic model and relative entropy. In Y. Chen, H. Deng, D. Zhang, and Y. Xiao (Eds.), *FSKD (5)*, pp. 384–387. IEEE Computer Society.

Yin, Z., X. Zhou, C. Bakal, F. Li, Y. Sun, N. Perrimon, and S. Wong (2008). Using iterative cluster merging with improved gap statistics to perform online phenotype discovery in the context of high-throughput RNAi screens. *BMC bioinformatics 9*(1), 264.

Zheng-Jun, Z. and Z. Yao-Qin (2009). Estimating the image segmentation number via the entropy gap statistic. In *ICIC '09: Proceedings of the 2009 Second International Conference on Information and Computing Science*, Washington, DC, USA, pp. 14–16. IEEE Computer Society.